\documentclass[onecolumn,showpacs,floatfix,aps,nofootinbib]{revtex4}
\usepackage{amssymb,amsmath}
\usepackage[dvips]{graphics,color}
\usepackage{epsfig}\usepackage{float}
\usepackage{natbib}
\usepackage{makeidx}
\usepackage{indentfirst}
\newcommand{\be}{\begin{equation}}
\newcommand{\ee}{\end{equation}}

\newcommand{\ba}{\begin{eqnarray}}
\newcommand{\ea}{\end{eqnarray}}

\begin{document}

\title{Gravitational constraints of $dS$ branes in $AdS$ Einstein-Brans-Dicke bulk}

\author{J. M. Hoff da Silva}
\email{hoff@feg.unesp.br} \affiliation{UNESP - Campus de Guaratinguet\'a - DFQ \\Av. Dr.
Ariberto Pereira da Cunha, 333\\ CEP 12516-410, Guaratinguet\'a - SP,
Brazil}

\author{Rold\~ao da Rocha}
\email{roldao.rocha@ufabc.edu.br} \affiliation{
Centro de Matem\'atica, Computa\c c\~ao e Cogni\c c\~ao,
Universidade Federal do ABC, 09210-170, Santo Andr\'e - SP, Brazil}


\begin{abstract}
We derive the full projected Einstein-Brans-Dicke gravitational equations associated with a
$n$-dimensional brane embedded in a $(n+1)$-dimensional bulk.
By making use of general conditions, as the positivity of the
Brans-Dicke parameter and the effective Newton
gravitational constant as well, we are able to constrain the brane
cosmological constant in terms of the brane tension, the
Brans-Dicke scalar field, and the trace of the stress tensor on the
brane, in order to achieve a $dS$ brane. Applying these constraints
to a specific five-dimensional model, a lower bound for
the scalar field on the brane is elicited without solving the full
equations. It is shown under which conditions the brane
effective cosmological constant can be ignored in the brane
projected gravitational field equations, suggesting a
different fine tuning between the brane tension and the bulk cosmological.
\end{abstract}
\maketitle

\section{Introduction}

Soon after the seminal papers of Randall and Sundrum
\cite{RSI,RSII}, the amount of works dealing with the theoretical
possibility of large extra dimensions in theories describing the Universe increased
significantly. The possibility of solving the hierarchy problem in
the Randall-Sundrum (RS) scenario, for instance, is one of the
major reasons for such an increase. There are even other
outstanding and prominent features arising in the scope of the gravitational
aspects related to RS-like models, accruing from a non-factorizable geometry and
$\mathbb{Z}_{2}$ symmetry. The basic setup concerning the RSI model formalism
\cite{RSI} is composed by two mirror domain walls --- 3-branes ---
as boundaries of a five-dimensional $AdS$ bulk. The extra dimension
is represented by the orbifold $S^{1}/\mathbb{Z}_{2}$. It is
shown in \cite{RSII} that the model is still consistent, if one of the
branes is taken to infinity.

Nevertheless, despite of all the RS models success, it may be
not the final word in braneworld models. In the light of the string
theory advances, there are at least two distinguished features that the
RS models cannot provide: models dealing with more than five dimensions and also other fields
mediating the gravitational phenomena. In fact, unification models
such as supergravity, superstrings, and $M$-theory \cite{GSW}
effectively predict the existence of a scalar gravitational field
acting as a gravitational interaction mediator, together
with the usual rank-2 tensor field. In this vein, it is
worthwhile to consider the gravitational field equations
in a scalar-tensor theory context. In this work we shall deal
with the simplest consistent scalar-tensor theory found in
the literature: the Brans-Dicke theory \cite{BD}.

There are several models concerning the accomplishment  of
braneworld scenarios, in many different dimensions. For instance,
going forward in the program consisting of the use of topological defects in order
to generate all the bulk/brane structure, the global cosmic string
was used in ordinary General Relativity \cite{CGE}, where
a six-dimensional spacetime is obtained. Moreover, the use of a local cosmic
string in Brans-Dicke gravity was scrutinized in \cite{PRIM}, while a
global cosmic string, again in Brans-Dicke theory, was analyzed in
\cite{SEG}. Besides, in General Relativity, this type of
six-dimensional scenario was investigated with an explicit local vortex
source \cite{14} and a blown-up  brane was analyzed in
\cite{15,16}, where a $\mathbb{Z}_{2}$ symmetry connects the
solutions inside and outside the brane.

A consistent mode to investigate the gravitational aspects and
effects about the presence of extra dimensions is to look at the
Gauss-Codazzi equations \cite{WALD}, resulting from a foliation
of the bulk spacetime and the projection of the gravitational field
equations from the bulk into the brane. It was, in fact, applied
to RS-like braneworlds in General Relativity in \cite{GCGR} with
$\mathbb{Z}_{2}$ symmetry. Such a procedure allows the investigation of a
wide variety of brane gravitational phenomena \cite{MAART}. Without the
$\mathbb{Z}_{2}$ symmetry, a similar but more involved procedure
was implemented in \cite{JAP}. In the context of Brans-Dicke
gravity, the Gauss-Codazzi formalism was implemented in a
six-dimensional braneworld with $\mathbb{Z}_{2}$ symmetry in
reference \cite{EP1}, and without such a symmetry in \cite{EP2}.
This dimensionality $(D=6)$ is maintained in such papers in order
to be consistent with the works aforementioned in the previous
paragraph. In the present paper we should leave this constraint
apart, working, instead, in arbitrary dimensions and scrutinizing a
five-dimensional example explicitly. A key characteristic for the
implementation of the Gauss-Codazzi formalism to braneworlds is
the codimension --- the number of extra dimensions {\it out} the
brane --- one. In the case of a codimension bigger than one it is
still possible to choose a direction to foliate the bulk,
nevertheless the projection of the bulk geometric quantities
cannot be implemented, since one needs the concept of different
``sides'' of the hypersurface that the equations are projected
on\footnote{There are important results concerning codimension two
braneworld models. In \cite{RG}, the addition of a Gauss-Bonnet
term in the Lagrangian allows the extraction of physical
information about the system. Recently, a rigorous approach to
codimension two was derived in \cite{GT}.}.

The aim of this paper is twofold: firstly,
 we are concerned to find out the full
projected equations from a $(n+1)$-dimensional bulk into a
$n$-dimensional brane with $\mathbb{Z}_{2}$ symmetry, in the
context of Brans-Dicke theory. It has been never accomplished before in the
Brans-Dicke theory and since it is the first step to extract some
physical information about such system, we shall generalize the
Gauss-Codazzi formalism to this case, for arbitrary dimensions.
Secondly and most important, by exploring the general form of the
resulting equations, we are able to provide some conductive formal constraints
which must be satisfied by a braneworld model in Brans-Dicke
gravity in arbitrary dimensions, in order to generate a specific
scenario as, for instance, a $dS$ brane embedded into an $AdS$ bulk.
After establishing the general picture, we scrutinize a concrete example of a 3-brane in a
five-dimensional bulk. It is shown that (for the vacuum on the brane) we are
able to find out a lower bound of the Brans-Dicke scalar field --- usually called
the ``dilaton'' hereon --- on the brane, {\it even without
solving} the full Einstein-Brans-Dicke equation. We also discuss
about the possibility of discarding the brane effective
cosmological constant term in the brane projected equations.

This paper is organized as follows: in the next Section the basic notations and conventions used in this
work are introduced. In Section III we indite the full projected
Einstein-Brans-Dicke equation on the brane for an arbitrary
dimension. Those two Sections are devoted to a brief revision
and also delve into a generalization of the Gauss-Codazzi formalism to an arbitrary bulk dimension,
in the scope of Brans-Dicke gravity. Thereafter, in Section IV
 the general constraints necessary to implement a
specific model are derived, namely a $dS$ brane embedded into a $AdS$ bulk. In addition, we analyze a concrete case by investigating a
five-dimensional RS-like braneworld. It constitutes
a prominent example of how the constraint obtained is useful in order to evince physical gravitational
information about the braneworld model dealt with. In the final
Section we conclude, summarizing our results, and point out
some future perspectives in this research line.

\section{Preliminaries}

This Section is devoted to recall the implementation of the
 Gauss-Codazzi formalism geometric part, starting from a
$(n+1)$-dimensional bulk. It was accomplished elsewhere \cite{FEITO} in
the General Relativity framework, but we shall keep this Section in
order to guarantee some sequential readability of the paper.
Moreover, the implementation of such formalism to the
Brans-Dicke case was accomplished only for specific scenarios, as the six-dimensional
case \cite{EP1,EP2}, but not regarding an arbitrary dimension. As
we shall see, this generalization provides a rigorous method to
explore some physical properties of the braneworld we are dealing with.

We start by regarding the brane as a $n$-dimensional time-like
hypersurface embedded in a $(n+1)$-dimensional
bulk, namely, a codimension one braneworld.
Denoting $n^{\mu}$ the components of an unitary vector orthogonal to the brane and
$g_{\mu\nu}$ the components of the bulk metric tensor of signature $(-,+,+,\ldots,+)$, the components of the
induced metric on the brane is given by
$q_{\mu\nu}=g_{\mu\nu}-n_{\mu}n_{\nu}$. Denoting the bulk covariant
derivative by $\nabla_{\mu}$, the Gauss equation forthwith reads \be
^{(n)}\!R^{\alpha}_{\;\,\beta\gamma\delta}=\,^{(n+1)}\!R^{\mu}_{\;\,\nu\rho\sigma}q^{\alpha}_{\mu}q^{\nu}_{\beta}q^{\delta}_{\gamma}q^{\sigma}_{\delta}+
K^{\alpha}_{\;\,\gamma}K_{\beta\delta}-K^{\alpha}_{\;\,\delta}K_{\beta\gamma}\label{1},
\ee where
$K_{\mu\nu}=q_{\mu}^{\alpha}q_{\nu}^{\beta}\nabla_{\alpha}n_{\beta}$
denotes the extrinsic curvature, indication the way how the brane is embedded
in the bulk. Equation (\ref{1}) basically asserts that the brane
curvature tensor is given by the projection of the bulk curvature
tensor and corrections coming from extrinsic curvature terms. The
Codazzi equation is given by \ba
D_{\nu}K^{\nu}_{\;\,\mu}-D_{\mu}K=\,^{(n+1)}\!R_{\rho\sigma}n^{\sigma}q^{\rho}_{\;\,\mu},
\label{2} \ea where $D_{\mu}$ denotes the covariant derivative on the
brane.
From Equation (\ref{1}) it is immediate to see that the $n$-dimensional
Ricci tensor is given by \ba
^{(n)}\!R_{\beta\delta}=\,^{(n+1)}\!R_{\nu\sigma}q^{\nu}_{\;\,\beta}q^{\sigma}_{\;\,\delta}-\,^{(n+1)}\!R^{\mu}_{\;\,\nu\rho\sigma}n_{\mu}n^{\rho}
q_{\;\,\beta}^{\nu}q_{\;\,\delta}^{\sigma}+KK_{\beta\delta}-K^{\;\,\gamma}_{\delta}K_{\beta\gamma},\label{3}
\ea while the scalar curvature is given by
\be
^{(n)}\!R=q^{\beta\delta}\,^{(n)}\!R_{\beta\delta}=q^{\beta\delta}\,^{(n+1)}\!R_{\beta\delta}-\,^{(n+1)}\!R^{\mu}_{\nu\rho\sigma}n_{\mu}
n^{\rho}q^{\nu\sigma}+K^{2}-K^{\alpha\beta}K_{\alpha\beta}.\label{4}
\ee Using Equations (\ref{2}) and (\ref{3}) it is possible to construct the
$n$-dimensional Einstein tensor $^{(n)}\!G_{\mu\nu}$: \ba
^{(n)}\!G_{\mu\nu}&=&\left.\,^{(n+1)}\!R_{\nu\sigma}q^{\nu}_{\beta}q^{\sigma}_{\delta}-\tilde{E}_{\beta\delta}+KK_{\beta\delta}-K^{\gamma}_{\delta}K_{\beta\gamma}
-\frac{1}{2}q_{\beta\delta}q^{\nu\sigma}\,^{(n+1)}\!R_{\nu\sigma}\right.\nonumber\\&& +\left.\frac{1}{2}q_{\beta\delta}q^{\nu\sigma}\,^{(n+1)}\!R^{\mu}_{\nu\rho\sigma}n_{\mu}n^{\rho}
q^{\nu\sigma}-\frac{1}{2}q_{\beta\delta}(K^{2}-K^{\alpha\gamma}K_{\alpha\gamma})\right.,\label{5}
\ea where
$\tilde{E}_{\beta\delta}=\,^{(n+1)}\!R^{\mu}_{\nu\rho\sigma}n_{\mu}n^{\rho}q^{\nu}_{\beta}q^{\sigma}_{\delta}$.
Now, taking into account the relations \ba
q_{\beta\delta}q^{\nu\sigma}\,^{(n+1)}\!R_{\nu\sigma}=g_{\nu\sigma}\,^{(n+1)}\!Rq^{\nu}_{\beta}q^{\sigma}_{\delta}-
q_{\beta\delta}\,^{(n+1)}\!R_{\nu\sigma}n^{\nu}n^{\sigma}\label{6}
\ea and \ba
^{(n+1)}\!R^{\mu}_{\;\,\nu\rho\sigma}n_{\mu}n^{\rho}q_{\nu\sigma}=\,^{(n+1)}\!R_{\mu\rho}n^{\mu}n^{\rho},\label{7}
\ea we arrive at \ba
^{(n)}\!G_{\beta\delta}&=&\left.\,^{(n+1)}\!G_{\nu\sigma}q_{\beta}^{\nu}q_{\delta}^{\sigma}+\,^{(n+1)}\!R_{\nu\sigma}n^{\nu}n^{\sigma}q_{\beta\delta}+
KK_{\beta\delta}-K_{\delta}^{\gamma}K_{\beta\gamma}\right.\nonumber\\&&-\left.
\frac{1}{2}q_{\beta\delta}(K^{2}-K^{\alpha\beta}K_{\alpha\beta})-\tilde{E}_{\beta\delta}.\right.\label{8}
\ea

It is useful to express the $n$-dimensional Einstein tensor in
terms of the bulk Weyl tensor. After all manipulations, the
Weyl tensor --- denoted hereon by $C^{\mu}_{\;\,\nu\rho\sigma}$ ---  brings some
genuine contributions from the bulk. The relation between
the Riemann, the Ricci, the Weyl tensors, and the scalar curvature
in $(n+1)$-dimensions is provided by \ba
^{(n+1)}\!R^{\mu}_{\;\,\nu\rho\sigma}=\,^{(n+1)}\!C^{\mu}_{\;\,\nu\rho\sigma}+\frac{2}{n-1}\Big(\,^{(n+1)}\!R^{\mu}_{[\rho}g_{\sigma]\nu}-
\,^{(n+1)}\!R_{\nu[\rho}g_{\sigma]}^{\mu}\Big)-\frac{2}{n(n-1)}\,^{(n+1)}\!Rg^{\mu}_{[\rho}g_{\sigma]\nu}.\label{9}
\ea The first important relation to notice is that
$\tilde{E}_{\beta\delta}$ can be expressed in terms of $E_{\beta\delta}\equiv
\,^{(n+1)}C^{\mu}_{\;\,\nu\rho\sigma}n_{\mu}n^{\rho}q^{\nu}_{\beta}q^{\sigma}_{\delta}$
by \ba
\tilde{E}_{\beta\delta}=E_{\beta\delta}+\frac{1}{n-1}\Big(\,^{(n+1)}\!R^{\mu}_{\rho}n_{\mu}n^{\rho}q_{\beta\delta}
+\,^{(n+1)}\!R_{\nu\sigma}q_{\beta}^{\nu}q_{\delta}^{\sigma}\Big)-\frac{1}{n(n-1)}\,^{(n+1)}\!Rq_{\beta\delta}.\label{10}
\ea Then, after some manipulations the brane Einstein
tensor can be expressed as the following: \ba
^{(n)}\!G_{\beta\delta}&=&\left.\frac{(n-3)}{(n-1)}\,\left(^{(n+1)}\!G_{\nu\rho}q_{\beta}^{\nu}q_{\delta}^{\rho}-\,^{(n+1)}\!
R_{\nu\sigma}q_{\beta\delta}q^{\nu\sigma}\right)+ \frac{(n^{2}-4n+2)}{n(n-1)}\,^{(n+1)}\!Rq_{\beta\delta}+KK_{\beta\delta}-
K^{\gamma}_{\delta}K_{\beta\gamma}\right.\nonumber\\&&\left.-\frac{1}{2}q_{\beta\delta}(K^{2}-K^{\alpha\gamma}K_{\alpha\gamma})-E_{\beta\delta}.\right.\label{11}
\ea The equation above encloses the main result of this Section.
In the next Section we express the geometric bulk quantities
--- $^{(n+1)}\!G_{\mu\nu}$, $^{(n+1)}\!R_{\mu\nu}$, and
$^{(n+1)}\!R$ --- in terms of the dilaton field and the
bulk energy-momentum tensor.

\section{The full projected Einstein-Brans-Dicke equation}

In order to extend the application of the Gauss-Codazzi formalism
to the Brans-Dicke theory for gravity in arbitrary dimensions, we start
by writing the $(n+1)$-dimensional Brans-Dicke equation \ba
^{(n+1)}G_{\nu\rho}=\frac{8\pi}{\phi}T_{\nu\rho}+\frac{w}{\phi^{2}}\Big(\nabla_{\nu}\phi\nabla_{\rho}\phi-\frac{1}{2}g_{\nu\rho}\nabla_{\alpha}\phi
\nabla^{\alpha}\phi\Big)+\frac{1}{\phi}\Big(\nabla_{\nu}\nabla_{\rho}\phi-g_{\nu\rho}\Box^{2}\phi\Big),\label{12}
\ea with \be \Box^{2}\phi=\frac{8\pi}{n+(n-1)w}T,\label{13}\ee where
$T_{\mu\nu}$ stands for the bulk energy-momentum tensor, encompassing and describing all the Universe content but the field $\phi$ and the gravity itself.

Substituting Equation (\ref{13}) in the last term of the
right-hand side of (\ref{12}), and contracting all the resulting
equation with $g^{\nu\rho}$ we have \be
^{(n+1)}\!R=\frac{-2w}{n+(n-1)w}\frac{8\pi}{\phi}T+\frac{w}{\phi^{2}}\nabla^{\alpha}
\phi\nabla_{\alpha}\phi.\label{14}\ee Returning to the Equation
(\ref{12}) the Ricci tensor is given by \ba
^{(n+1)}\!R_{\nu\rho}=\frac{8\pi}{\phi}T_{\nu\rho}+\frac{w}{\phi^{2}}\nabla_{\nu}\phi\nabla_{\rho}\phi+\frac{1}{\phi}\nabla_{\nu}\nabla_{\rho}\phi-
\frac{8\pi}{\phi}g_{\nu\rho}\frac{w+1}{n+(n-1)w}T.\label{15}
\ea Now, the first three terms on the
right-hand side of Eq.(\ref{11}) can be computed. The resulting equation expressed in terms of the
dilaton field and the stress-tensor is given by \ba
^{(n)}\!G_{\beta\delta}&=&\left.\frac{(n-3)}{(n-1)}\left(\frac{8\pi}{\phi}T_{\nu\rho}+\frac{w}{\phi^{2}}\nabla_{\nu}\phi\nabla_{\rho}\phi+
\frac{1}{\phi}\nabla_{\nu}\nabla_{\rho}\phi\right)(q^{\nu}_{\beta}q^{\rho}_{\delta}-q_{\beta\delta}q^{\nu\rho})+\frac{(n-4)}{2n}\;
\frac{w}{\phi^{2}}q_{\beta\delta}\nabla_{\alpha}\phi\nabla^{\alpha}\phi\right.\nonumber\\&&-\left.\frac{8\pi}{\phi}\frac{T}{[n+(n-1)w]}q_{\beta\delta}
\Big[(w+1)(1-n)+\frac{(n-4)w}{n}\Big] + KK_{\beta\delta}+
K^{\gamma}_{\delta}K_{\beta\gamma}\right.\nonumber\\&&-\left.\frac{1}{2}q_{\beta\delta}(K^{2}-K^{\alpha\beta}K_{\alpha\beta})-E_{\beta\delta}.\right.\label{16}
\ea On the other hand, from Equations  (\ref{2}) and (\ref{15}) it follows that \be
D_{\nu}K^{\nu}_{\;\,\mu}-D_{\mu}K=\left(\frac{8\pi}{\phi}T_{\nu\rho}+\frac{w}{\phi^{2}}\nabla_{\nu}\phi\nabla_{\rho}\phi+
\frac{1}{\phi}\nabla_{\nu}\nabla_{\rho}\phi\right)n^{\rho}q_{\mu}^{\nu}.\label{17}
\ee

Hereon in this Section our efforts will be focused to
effectively project Equation (\ref{16}) on the brane. It is
accomplished by the generalization of the Israel-Darmois \cite{ID}
junction condition to the Brans-Dicke case. By using standard
tools of the distributional calculus, it is shown in the Appendix
of reference \cite{EP1} that the generalized junction condition
reads (for a $(n+1)-$dimensional bulk) \be
[K_{\mu\nu}]-[K]q_{\mu\nu}=-\frac{8\pi}{\phi}\Big(T^{brane}_{\mu\nu}-q_{\mu\nu}\frac{T^{brane}}{n+(n-1)w}\Big),\label{18}
\ee where the quantity $[X]$ means $[X]=X^{+}-X^{-}$. Here $X^{\pm}$ denotes the projection of any tensor, $X$, into the
brane by the $\pm$ side. We refer the reader to the reference
\cite{EP1} for all the details. Apart from that, we stress that
$T^{brane}_{\mu\nu}$ is the energy-momentum tensor on the brane
and $T^{brane}$ its respective trace. As we will see, these terms
bring contribution from the brane tension and the matter on
the brane as well. From the trace of Equation (\ref{18}) it is straightforward to see
that \be
[K]=\frac{8\pi}{\phi}T^{brane}\frac{w}{n+(n-1)w}.\label{19}
\ee Then, returning to (\ref{18}) one finds \ba
[K_{\mu\nu}]=-\frac{8\pi}{\phi}T^{brane}_{\mu\nu}+\frac{8\pi}{\phi}\frac{q_{\mu\nu}(w+1)T^{brane}}{n+(n-1)w}.\label{20}
\ea Now, the quantities in brackets can be forthwith determined by the
imposition of the $\mathbb{Z}_{2}$ symmetry. The role of the
$\mathbb{Z}_{2}$ symmetry in braneworld models is multiple
\cite{SUND}, but in what concerns its immediate effect on the
gravitational equations, it turns out to be quite straightforward: it just
changes the sign of the unitary orthogonal vector field, $n^{\alpha}$,
across the brane. Imposing such a symmetry, it implies that $n_{\alpha}^{+}\mapsto -n_{\alpha}^{-}$. From the
 extrinsic curvature definition, one concludes that the same
effect holds for $K_{\mu\nu}$, namely
$K_{\alpha\beta}^{+}\mapsto -K_{\alpha\beta}^{-}$. This result
enables us to write Equations (\ref{19}) and (\ref{20}) as
the following: \ba
K^{+}=\frac{4\pi}{\phi}T^{brane}\frac{w}{n+(n-1)w}\label{21},\qquad\qquad\qquad
K_{\mu\nu}^{+}=\frac{4\pi}{\phi}T^{brane}_{\mu\nu}+\frac{4\pi}{\phi}\frac{(w+1)q_{\mu\nu}T^{brane}}{n+(n-1)w},\label{22}
\ea respectively. Note that in Equation (\ref{16}) the terms
of extrinsic curvature are quadratic. Therefore, the values of $K$ and $K_{\mu\nu}$ can be computed at any side of the brane, in such way that the label $\pm$ can be concealed without any detriment in the theory.

Substituting the equations above into the Equation (\ref{16}),
suppressing the labels $\pm$, it follows that \ba
^{(n)}\!G_{\beta\delta}&=&\left.\frac{(n-3)}{(n-1)}\left(\frac{8\pi}{\phi}T_{\nu\rho}+\frac{w}{\phi^{2}}\nabla_{\nu}\phi\nabla_{\rho}\phi
+\frac{1}{\phi}\nabla_{\nu}\nabla_{\rho}\phi\right)(q_{\beta}^{\nu}q_{\delta}^{\rho}-q_{\beta\delta}q^{\nu\rho})+\frac{(n-4)}{2n}
\frac{w}{\phi^{2}}q_{\beta\delta}\nabla_{\alpha}\phi\nabla^{\alpha}\phi\right.\nonumber\\&&-\left.\frac{8\pi}{\phi}\frac{q_{\beta\delta}T}
{n+(n-1)w}\Big[(w+1)(1-n)+\frac{(n-4)w}{n}\Big]+\Bigg(\frac{16\pi^{2}}{\phi^{2}}\Bigg)\Bigg\{\frac{1}{2}q_{\beta\delta}
T^{brane}\,^{\alpha\gamma}T^{brane}_{\alpha\gamma}\right.\nonumber\\&-&\left. T^{brane}_{\;\delta}\,^{\gamma}T^{brane}_{\beta\gamma}-\frac{(w+2)}{n+(n-1)w}T^{brane}T^{brane}_{\beta\delta}
+q_{\beta\delta}(T^{brane})\,^{2}\,\frac{[(w+1)^{2}(3n-2)-w^{2}]}{2[n+(n-1)w]^{2}}\Bigg\}-E_{\beta\delta}\right.\label{23}. \ea

We remark that the quantities involved in Equation (\ref{23})
are interpreted in the limit of the extra transverse dimension
approaching the brane. Equation (\ref{23}) is written in a far
from suitable way, in order to distinguish between the two stress-tensors, i. e. the two distinct
quantities playing the role of sources, $T_{\mu\nu}$ and $T^{brane}_{\mu\nu}$. As a matter of fact,
$T_{\mu\nu} \supset T^{brane}_{\mu\nu}$ and to complete the analysis we have to specify the form of the
stress-tensors. Ignoring any type of bulk source, except a
cosmological constant and the brane itself, one can write the bulk
stress tensor as \ba T_{\mu\nu}=-\Lambda
g_{\mu\nu}+\delta(y-y_{b})T^{brane}_{\mu\nu},\label{25}\ea where
$\delta(y)$ is necessary in order to position the brane
(generically assumed in $y=y_{b}$) in the bulk, and \ba
T^{brane}_{\mu\nu}=-\lambda q_{\mu\nu}+\tau_{\mu\nu},\label{26}
\ea where $\lambda$ denotes the brane tension and $\tau_{\mu\nu}$ the
contribution of any matter field to the brane energy-momentum
tensor. Substituting Equations (\ref{25}) and (\ref{26}) into
(\ref{23}), and taking into account that all the
quantities are computed in the limit approaching the brane, the
following result accrues:\ba
^{(n)}\!G_{\beta\delta}&=&\left.\frac{(n-3)}{(n-1)}\Bigg(\frac{w}{\phi^{2}}\nabla_{\nu}\phi\nabla_{\rho}\phi
+\frac{1}{\phi}\nabla_{\nu}\nabla_{\rho}\phi\Bigg)(q_{\beta}^{\nu}q_{\delta}^{\rho}-q_{\beta\delta}q^{\nu\rho})-
\Lambda_{n}q_{\beta\delta}+8\Omega\tau_{\beta\delta}+16\pi^{2}\Sigma_{\beta\delta}-E_{\beta\delta}\right.,\label{27}
\ea where $\Omega$, $\Lambda_{n}$, and $\Sigma_{\beta\delta}$ are
given, respectively, by \ba\Omega=\frac{2\pi\lambda}{\phi^{2}}\frac{[4n+(3n-2)w]}{n+(n-1)w}\label{28},\ea
\ba \Lambda_{n}&=&\left.\frac{8\pi}{\phi}\frac{(-\Lambda)[(1-3n)(w+1)-4w]}{n[n+(n-1)w]}-\frac{(n-4)}{2n}\frac{w}{\phi^{2}} \nabla^{\alpha}\phi\nabla_{\alpha}\phi-\frac{16\pi^{2}}{\phi^{2}}\frac{\lambda^{2}}{[n+(n-1)w]^{2}}\right.\nonumber
\\&\times&\left.\Big\{(n-1)(3n^{2}+n+2)w^{2}+n(8n^{2}-5n+2)w+n^{2}(5n-2)\Big\}-\frac{16\pi^{2}}{\phi^{2}}
\frac{\lambda\tau}{[n+(n-1)w]^{2}}\right.\nonumber\\&\times&\left. \Big\{nw^{2}+(w+1)[4n(1-n)+w(-4n^{2}+5n-2)]\Big\} \right.\label{30}
\ea
and
\ba\hspace{-0.7cm}\Sigma_{\beta\delta}&=&\left.\frac{1}{\phi^{2}}\Bigg(\frac{[(w+1)^{2}(3n-2)-w^{2}]}{2[n+(n-1)w]^{2}}
q_{\beta\delta}\tau^{2}+\frac{1}{2}q_{\beta\delta}\tau^{\alpha\gamma}\tau_{\alpha\gamma}
-\Bigg(\frac{w+2}{n+(n-1)w}\Bigg)\tau\tau_{\beta\delta}-\tau_{\delta}^{\gamma}\tau_{\beta\gamma}\Bigg)\right.
\label{31} \ea

We shall conclude this Section providing some useful interpretations about
the results encoded in Equations (\ref{27})--(\ref{31})
\cite{GCGR,EP1}. We write the full projected equation in an
Einstein tensor-like form. The first term of (\ref{27}) brings
the specific contribution of the scalar field dynamics, and it must be carefully
analyzed in any cosmological application of Equation (\ref{27}).
In fact, we expect that such a term play an important role in cosmological scenarios. The second
term is given by the effective brane cosmological constant,
$\Lambda_{n}$, which depends on the bulk cosmological constant,
the brane tension, the dilaton field, and on the stress-tensor
trace of matter on the brane. We shall analyze carefully this term in the
next Section. The term proportional to
$\tau_{\mu\nu}$ is analogue to the usual source term of Einstein
equation, where $\Omega$ is the effective Newtonian constant. The
penultimate term is quadratic on the brane stress-tensor and could
play a very important role in early stages of the cosmological
evolution. The last term proportional to the Weyl tensor brings
genuine contribution coming from the bulk, and has no analogue in
the usual four-dimensional case.

Note that, from Equation (\ref{28}), the sign of the
Newtonian effective gravitational constant strongly depends on the
sign of the brane tension $\lambda$. It will be our starting point
to derive the constraints in the next Section. As we will see,
the functional form of the scalar field plays an important role
in the dependence of $\Omega$ on the brane tension. Before beginning
such analysis, however, we want to comment two more remarks about
$\Omega$. First, its dependence on $\lambda$ tell us that it could
not be possible to define gravity in some cosmological era before
the formation of structures. Apart from that, the dependence on
$1/\phi^{2}$ turns explicit the importance of the dilaton field stabilization,
in order to guarantee the agreement with usual
gravity on the brane. In some artificial way, this situation can
be used to fix the brane position along the transverse extra
dimension, leading then to the right value to the scalar field. In
a more rigorous way, the stabilization of the dilaton field can
be accomplished, for instance, by the introduction of a well behaved
potential in the Brans-Dicke part of the action \cite{PERIVO}.

\section{General constraints: A five-dimensional example}

In this Section we shall derive some general constraints relating
the bulk cosmological constant, the brane tension, and the scalar
field, in order to lead to a specific scenario, namely, a $dS$ brane
$(\Lambda_{n}\gtrsim 0)$, accordingly to the low expansion
observations, with a positive Newton's effective gravitational
constant \cite{CGK} embedded into an $AdS$ bulk.

Before proceeding, let us state a relevant remark about Equation
(\ref{28}). Note that the brane tension must be positive in order to
engender the right sign to the effective gravitational constant. It is
in accordance with the results previously obtained in the literature \cite{POS,REFER},
and is also consistent with a physical gravitational object, since a negative tension
brane is intrinsically unstable. It is purposeful to remark that for certain compactification models, the
Brans-Dicke parameter may be negative. In this case, it is still possible
to have a positive brane tension and $\Omega >0$, provided that $|w|<3/(2-1/n)$. Henceforward we restrict the analysis for positive Brans-Dicke parameter.

Now, we focus on a scenario consisting of an $AdS$ ($\Lambda <0$) bulk
with an embedded $dS$ ($\Lambda_{n}\gtrsim 0$) brane.
As we shall see, the functional form for the dilaton field on the brane can be extracted, providing
an important boundary condition of such scalar degree of freedom. By looking
at the explicit form of the effective cosmological constant (\ref{30}) and imposing the
above conditions, it implies --- for a non-vanishing scalar field --- that
\begin{eqnarray}&&
\left. \frac{|\Lambda|}{\phi}\frac{[(1-3n)(w+1)-4w]}{n[n+(n-1)w]}\gtrsim \frac{(n-4)}{2n}\frac{w}{8\pi\phi^{2}}\nabla^{\alpha}\phi\nabla_{\alpha}\phi+\frac{2\pi}{\phi^{2}}
\frac{\lambda^{2}}{[n+(n-1)w]^{2}}\right.\nonumber\\&\times&\left. \{(n-1)(3n^{2}+n+2)w^{2}+n(8n^{2}-5n+2)w+n^{2}(5n-2)\}+\frac{2\pi}{\phi^{2}}\frac{\lambda\tau}{[n+(n-1)w]^{2}}
\right.\nonumber\\&\times&\left.\{nw^{2}+(w+1)[4n(1-n)+w(-4n^{2}+5n-2)]\}.\label{volta1}\right.
\end{eqnarray}

The equation above is written in a form that
completely obscures the associated physical content. Nevertheless, it may be useful by the
specification of a particular scenario. Let us investigate the
example of a RS-like model, composed by a 3-brane $(n=4)$ embedded in
the bulk. Note that in such dimensionality there is
a great simplification, since the term $\nabla_{\alpha}\phi\nabla^{\alpha}\phi$
appearing in (\ref{volta1}) vanishes. Just by investigating
this general constraint, it is obtained a lower bound for the dilaton field, in terms of the bulk cosmological constant, the brane tension and the Brans-Dicke parameter, without solving the full projected equation. Going further, we show what type of relationship between the bulk cosmological
constant and the brane tension is necessary for $\Lambda_{4}$ be discarded of the projected
effective brane gravitational equation. Comparing with the case in General Relativity, this sufficient condition is a genuine output of the Brans-Dicke bulk gravity. As will be shown, this type of interplay
between the bulk cosmological constant and the brane tension is
intrinsically different from the usual RS fine tuning.

Let us make a brief comment about the approach outlined in the previous paragraph. The basic idea is to derive gravitational constraints, which provides physical information about the system encoded in the Equations (\ref{27})--(\ref{31}). Here, two points should be stressed. Firstly, the generality of the Gauss-Codazzi procedure rests upon the fact that we do not need to make any consideration about the specific functional form of the metric. In this vein, the derivation of simple, and informative, gravitational constraints without solving the projected equation appears to be an interesting approach. Besides, the solution of the ($\phi$, $q_{\mu\nu}$)-system may be far from trivial even in the simplest case. Let us analyze an short example in order to make more clear this second point. Suppose the dilaton depending only on the extra transverse dimension. In view of the Equation (\ref{13}), it is clear that $\Box^{2}\phi=\alpha$, being $\alpha$ a constant involving the brane tension and the bulk cosmological constant. This scenario may be achieved by the consideration of the brane vacuum, for instance. So, the resulting differential equation reads
\be \frac{d^{2}\phi}{dy^{2}}+f(x^{\mu})\frac{d\phi}{dy}=\alpha\label{vai1},\ee where $f(x^{\mu})$ is given by $f(x^{\mu})=\frac{1}{2}g^{\mu\nu}\partial_{y}g_{\mu\nu}$ and the bulk metric is diagonal. The solution for the Equation (\ref{vai1}) may be settled in the form \be \phi=\int \Bigg\{\alpha\int \exp{\Bigg(\int f(x^{\mu}) dy\Bigg)}+C_{1}\Bigg\}\exp{\Bigg(-\int f(x^{\mu}) dy\Bigg)}dy+C_{2},\label{vai2}\ee being $C_{1,2}$ constants. Even being the $g_{yy}$ component trivial, the brane metric may still have a complicated dependence on the extra dimension. It is, of course, the entire point of non-factorizable (warped) geometries of which a RS-like model is a simple example. Hence, the integration of $f(x^{\mu})$ may not be trivial. Apart from that, the robustness of this approach is, as remarked, in the generality of the unknown metric. Thus, we shall not to restrict the argumentation by specifying the brane metric.

Now, going further in our example, it is immediate to note that for $n=4$, Equation (\ref{volta1}) reads
\ba &&
\frac{-|\Lambda|}{\phi}\gtrsim \frac{2\pi}{\phi^{2}}\frac{8\lambda}{(4+3w)(11+15w)}\Big\{4(21w^{2}+55w+36)\lambda-
(21w^{2}+47w+24)\tau\Big\}\label{volta2}
\ea We shall derive a result comparable with the standard procedure developed in the General Relativity realm. In this vein we turn our attention to the brane vacuum \cite{MAART}. Note, however, that it appears inconsistent for a positive dilaton field on the brane, since the resulting constraint is $-|\Lambda|\gtrsim Q$, being $Q$ a strictly positive quantity. So, one are led to consider a negative dilaton field. In this vein, for a vacuum on the brane it is straightforward to derive, from (\ref{volta2}), a lower bound for the dilaton field projected on the brane
\ba
|\phi|\gtrsim\frac{64\pi \lambda^{2}}{|\Lambda|}\frac{(21w^{2}+55w+36)}{(45w^{2}+93w+44)}.\label{volta3}
\ea Equation (\ref{volta3}) asserts relevant information that shall be remarked. The general form $|\phi|\thicksim\frac{\pi\lambda^{2}}{|\Lambda|}$, is quite simple, nevertheless it is far from trivial. Again, the scalar field obviously is completely determined by the full solution of the bulk field equations, which may be a very difficult task. Instead, by simple inputs (necessary for the fixation of a particular scenario, but keeping considerable generality\footnote{Note that we do not assume any form for the metric.}) it is possible
to derive a constraint that gives the boundary value for the scalar field. We remark that,
from Equation (\ref{volta3}), it is possible to achieve an upper bound for the effective Newtonian constant on the brane, as well as its correct dependence on the brane tension in such scalar-tensor gravity scenario.
In fact, taking into account Equation (\ref{28}) for $n=4$ it follows that
\ba
\Omega\lesssim \frac{|\Lambda|^{2}}{\lambda^{3}}\Delta(w),\label{volta4}
\ea where $\Delta(w)=\frac{(15w^{2}+56w+48)(11+15w)^{2}}{[32(21w^{2}+55w+36)]^{2}}$.
The Equation (\ref{volta4}) stress once again the impossibility about the definition of gravity before
the structure formation. Note also, that it keeps the dependence of $\Omega$ on the $\lambda$ signal.

Now, it is possible to further explore the condition making the 
approximation $\Lambda_{4}\sim 0$ valid. This approximation
is widely used in the application of braneworld
models to cosmological problems as, for instance, the fitting of
the galactic rotation curves without dark matter and corrections
in the black hole area in General Relativity \cite{MAART,COSMO,ROLDAO/CARLAO}.
The result encoded in the Equation (\ref{volta3}) revels that if the dilaton obeys such a constraint, then the desired behavior for the projected cosmological constant is achieved $(\Lambda_{4}\gtrsim 0)$. It is immediate to see that for a huge bulk cosmological constant in comparing to the brane tension, namely $|\Lambda|\gg \lambda^{2}$, the constraint (\ref{volta3}) is easily satisfied. Note that it is not so trivial
as it may sound. If we compare to the four-dimensional case in
usual General Relativity, the effective brane cosmological
constant $(\Lambda_{GR4})$ is given by \ba
\Lambda_{GR4}=\frac{\kappa_{5}^{2}}{2}\Bigg(\Lambda+\frac{1}{6}\kappa_{5}^{2}\lambda^{2}\Bigg),\label{63}
\ea where $\kappa_{5}$ is the five-dimensional gravitational
coupling. So, the imposition of $|\Lambda|\gg \lambda^{2}$ in Equation
(\ref{63}) is not enough to guarantee the approximation
$^{(4)}\!G_{\mu\nu}=-E_{\mu\nu}$, since there is another term,
proportional to the brane tension itself. This is closely related to the
fine tuning between the bulk cosmological constant and the brane
tension of the RS model. These two quantities must be of the same order for
the cancellation of the $\Lambda_{4}$ contribution. In the
Brans-Dicke gravity context, however, the dynamics of the scalar
field --- the analogue to $1/\kappa_{5}^{2}$
--- enables us to bound both terms of $\Lambda_{4}$ in an unique term and,
since the behavior of the dilaton field obeys $\phi\sim
1/|\Lambda|$, it is quite enough a huge bulk cosmological
constant. 

It is important to stress, besides, another appreciable departure of the Einstein-Brans-Dicke case to the usual scenario in General Relativity. Going further in the vacuum on the brane case, Equation (\ref{27})
forthwith reads \ba
^{(4)}\!G_{\beta\delta}=\frac{1}{3}\Bigg(\frac{w}{\phi^{2}}\partial_{\nu}\phi\partial_{\rho}\phi+\frac{1}{\phi}\nabla_{\nu}
\partial_{\rho}\phi\Bigg)(q_{\beta}^{\nu}q_{\delta}^{\rho}-q_{\beta\delta}q^{\nu\rho})-\Lambda_{4}q_{\beta\delta}-E_{\beta\delta}.\label{54}
\ea
In this case then, even discarding the effective cosmological
constant, we shall expect some subtle but substantial modifications
coming genuinely from the scalar field dynamics in the analysis of
specific cosmological systems, specially in the context of a weak dilaton field on the brane. In other words, the first term of the Equation (\ref{54}) can provide a
good laboratory for the study of General Relativity departures in
the braneworld context. 

As an aside remark, we call some attention to fact that the first two terms of the Equation (\ref{54}) may obey a specific balance in order to provide a null brane scalar of curvature. From (\ref{54}) we have
\ba ^{(4)}\!R=\Bigg(\frac{w}{\phi^{2}}\partial_{\nu}\phi\partial_{\rho}\phi+\frac{1}{\phi}\nabla_{\nu}
\partial_{\rho}\phi\Bigg)q^{\nu\rho}+4\Lambda_{4}.\label{taca}\ea Therefore, this two terms may cancel out one each other resulting in a large scale flat brane. This possibility, however, may preclude a departure of Brans-Dicke brane scenarios from the usual General Relativity ones, and, obviously, shall be based upon some dynamical mechanism. Perhaps, an similar analysis to the one carried out in reference \cite{DAMNOR} shall shed some light in such a mechanism. In any case, the possibility raised in this paragraph needs further study.

\section{Final remarks and outlook}

Before weaving some final remarks we shall briefly summarize the
main procedures and results of this paper. Motivated by string
theory low energy recovered gravity, we started finding out the full
projected Einstein-Brans-Dicke equation in an arbitrary dimension
via the application of the Gauss-Codazzi mechanism. Up to our knowledge, it was never accomplished before, for arbitrary dimensions,
to the Brans-Dicke gravity case. After that, we found the necessary condition --- relating the
scalar field, the bulk cosmological constant, and the brane tension
--- under which we obtain a $dS$ brane embedded in an $AdS$ bulk.
Applying our results to a concrete five-dimensional model we are
able to set the profile of the dilaton field on a brane vacuum.

Using the profile of the dilaton field on the brane we
derived the general form of the effective brane cosmological constant,
arriving at the sufficient
condition $(|\Lambda|\gg \lambda^{2})$ that makes possible to discard the
$\Lambda_{4}$ term of the projected
gravitational equation. This fact is a genuine output of the Brans-Dicke gravity.
It indicates a different type of fine tuning between the bulk cosmological constant and the
brane tension in the Brans-Dicke gravity framework.

As a remark aside, we stress that in our analysis, we did not treated extensively the possibility of a negative
Brans-Dicke parameter. It is well known that to some specific
compactification models, the interplay between string theory low
energy gravity and Brans-Dicke theory is given by a negative $w$. However,
by keeping the usual Brans-Dicke gravity motivation we restrict our analysis
to the positive $w$ case.

This work is a first step in order to investigate and to delve into some cosmological
properties of Brans-Dicke braneworld models. As our results
indicates, it is expected subtle but important departures of those
cases analyzed in the scope of General Relativity. In this
context, specific problems as the variation of quasar luminosity
and corrections on the black holes areas \cite{ROLDAO/CARLAO}, just to enumerate some physically
interesting and relevant systems, which are normally investigated due the presence of extra dimensions, deserve
more attention in the context of Brans-Dicke braneworld gravity.

\section*{ACKNOWLEDGMENTS}

J. M. Hoff da Silva thanks FAPESP (PDJ 2009/01246-8) and Universidade Federal do ABC
(UFABC) where this work was partially accomplished. R da
Rocha thanks CNPq 304862/2009-6 for financial support.

\end{document}